\documentclass[prd,superscriptaddress,showpacs,showkeys,amsmath,amssymb]{revtex4}
\usepackage{slashed,graphicx,color,axodraw,hyperref}
\newtheorem{theorem}{Theorem}
\newtheorem{proposition}{Proposition}

\begin{document}
\title{Trace Formulae of Characteristic Polynomial and Cayley-Hamilton's Theorem, and
Applications to Chiral Perturbation Theory and General
Relativity}
\author{Hong-Hao Zhang}
\affiliation{%
Department of Physics, Tsinghua University, Beijing 100084, China}

\author{Wen-Bin Yan}
\affiliation{%
Department of Physics, Tsinghua University, Beijing 100084, China}

\author{Xue-Song Li}
\affiliation{%
Science College, Hunan Agricultural University, Changsha 410128,
China}

\begin{abstract}
By using combinatorics, we give a new proof for the recurrence
relations of the characteristic polynomial coefficients, and then we
obtain an explicit expression for the generic term of the
coefficient sequence, which yields the trace formulae of the
Cayley-Hamilton's theorem with all coefficients explicitly given,
and which implies a byproduct, a complete expression for the
determinant of any finite-dimensional matrix in terms of the traces
of its successive powers. And we discuss some of their applications
to chiral perturbation theory and general relativity.
\end{abstract}\pacs{02.10.Ud, 02.10.Ox, 12.39.Fe, 04.20.-q.}
\keywords{characteristic polynomial coefficients, Cayley-Hamilton's
theorem, chiral perturbation theory, general relativity.} \maketitle
\section{Introduction}

There is a famous theorem named in honor of Arthur Cayley and
William Hamilton in linear algebra, which asserts that any $n\times
n$ matrix $\mathbf{A}$ is a solution of its associated
characteristic polynomial $\chi_{\mathbf{A}}$ \cite{VVV:1983}. In
the popular $SU(2)\times SU(2)$ and $SU(3)\times SU(3)$ chiral
perturbation theories, the Cayley-Hamilton theorem has been used to
eliminate redundant terms and keep only independent pieces of the chiral
Lagrangian
\cite{Urech:1994hd,Fearing:1994ga,Bijnens:1999sh,Scherer:2002tk}.
The trace formulae of the Cayley-Hamilton theorem for $2\times2$ and
$3\times3$ matrices are respectively
\begin{eqnarray}
&&\mathbf{A}^2-\mathrm{tr}(\mathbf{A})\mathbf{A}+\frac{1}{2}[\mathrm{tr}^2(\mathbf{A})
-\mathrm{tr}(\mathbf{A}^2)]=\mathbf{0}\;,\\
&&\mathbf{A}^3-\mathrm{tr}(\mathbf{A})\mathbf{A}^2+\frac{1}{2}[\mathrm{tr}^2(\mathbf{A})
-\mathrm{tr}(\mathbf{A}^2)]\mathbf{A}-\frac{1}{6}[\mathrm{tr}^3(\mathbf{A})
-3\mathrm{tr}(\mathbf{A}^2)\mathrm{tr}(\mathbf{A})+2\mathrm{tr}(\mathbf{A}^3)]
=\mathbf{0}\;.
\end{eqnarray}
And a recursive algorithm to compute the characteristic polynomial
coefficients of a generic $n\times n$ matrix as functions of the
traces of its successive powers was given by Silva nearly
ten years ago \cite{Silva:98}. The work of Ref. \cite{Silva:98}
obtained the recurrence relations for
the coefficient sequence by using an improved version of the
remainder theorem and some additional results.

In this paper, with the knowledge of combinatorics, we give a new
proof for these recurrence relations, and then we further obtain an
explicit formula for the generic term of the coefficient sequence,
which yields the trace formulae of Cayley-Hamilton's theorem with
all coefficients explicitly given, and which implies a byproduct, a
complete expression for the determinant of any finite-dimensional
matrix in terms of traces of its successive powers. We also discuss
some applications to chiral perturbation theory and general
relativity.

This paper is organized as follows. In Sec.
\ref{sec-trace-formulae}, by using combinatorics, a new proof for
the recurrence relations of the characteristic polynomial
coefficients is given and the explicit expression for the generic
term of the coefficient sequence is obtained, which improves the
results of previous works. In Sec. \ref{sec-app-chpt}, a detailed
discussion of the applications of the Cayley-Hamilton theorem to
$n$-flavor chiral perturbation theories up to $n=5$ is presented. In
Sec. \ref{sec-gr}, as a further application, the determinant of the
metric tensor in 10 dimensional spacetime is explicitly expressed by
the traces of its successive powers. Sec. \ref{sec-con} is devoted
to conclusions. A check for the Cayley-Hamilton relations using an
alternative approach is presented in App. \ref{check}, while a {\it
PASCAL} code for the computation of the generic term of the
coefficient sequence is given in App. \ref{pascal-code}.

\section{Trace Formulae\label{sec-trace-formulae}}

Let $\mathbf{A}$ be any complex $n\times n$ matrix with eigenvalues
$\lambda_1$, $\lambda_2$, $\cdots$, $\lambda_n$ (possibly complex
and identical). Then its trace is
$\mathrm{tr}(\mathbf{A})=\lambda_1+\lambda_2+\cdots+\lambda_n$, and
the characteristic polynomial is
$\chi_{\mathbf{A}}(\lambda)=\det(\lambda
\mathbf{I}-\mathbf{A})=(\lambda-\lambda_1)(\lambda-\lambda_2)\cdots(\lambda-\lambda_n)$,
where $\mathbf{I}$ is the $n\times n$ unit matrix. And $\mathbf{A}$
being similar to a Jordan normal form implies that
$\mathrm{tr}(\mathbf{A}^m)=\lambda_1^m+\lambda_2^m+\cdots+\lambda_n^m$
for any positive integer $m$. Following the notation of Ref.
\cite{Silva:98}, the traces are denoted by
$T_m\equiv\mathrm{tr}(\mathbf{A}^m)$ here and henceforth. Now we are
ready to give a new proof for the theorem relating the coefficients
of $\chi_{\mathbf{A}}(\lambda)$ to the traces of $\mathbf{A}^m$
($m=1,2,\cdots,n$).

\begin{theorem}\label{theorem1}
If
$\chi_{\mathbf{A}}(\lambda)=\lambda^n+D_1\lambda^{n-1}+\cdots+D_{n-1}\lambda+D_n$
is the characteristic polynomial of the complex $n\times n$ matrix
$\mathbf{A}$ and $T_m=\mathrm{tr}(\mathbf{A}^m)$, then
\begin{eqnarray}
mD_m+D_{m-1}T_1+D_{m-2}T_2+\cdots+D_1T_{m-1}+T_m=0\;,\qquad
m=1,2,\cdots,n\;.
\end{eqnarray}
\end{theorem}

{\bfseries Proof.} Let $\lambda_1$, $\lambda_2$, $\cdots$,
$\lambda_n$ be the $n$ eigenvalues of the matrix $\mathbf{A}$. Then
$\chi_{\mathbf{A}}(\lambda)=\lambda^n+D_1\lambda^{n-1}+\cdots+D_{n-1}\lambda+D_n
=(\lambda-\lambda_1)(\lambda-\lambda_2)\cdots(\lambda-\lambda_n)$
yields
\begin{eqnarray*}
D_m=(-)^m\sigma_m\;,\qquad
\sigma_m\equiv\sum_{i_1<i_2<\cdots<i_m}\lambda_{i_1}\lambda_{i_2}\cdots\lambda_{i_m}\;,
\qquad m=1,2,\cdots,n\;.
\end{eqnarray*}
Multiplying $\sigma_{m-1}$ by
$(\lambda_1+\lambda_2+\cdots+\lambda_n)$, we have
\begin{eqnarray*}
\sigma_{m-1}T_1&=&(\lambda_1+\lambda_2+\cdots+\lambda_n)\cdot
\sum_{i_1<i_2<\cdots<i_{m-1}}\lambda_{i_1}\lambda_{i_2}\cdots\lambda_{i_{m-1}}\nonumber\\
&=&\sum_{i=1}^n\lambda_i^2\cdot\sum_{\substack{j_1<j_2<\cdots<j_{m-2}\\j_l\neq
i}} \lambda_{j_1}\lambda_{j_2}\cdots\lambda_{j_{m-2}}
+\sum_{i=1}^n\lambda_i\cdot\sum_{\substack{j_1<j_2<\cdots<j_{m-1}\\j_l\neq
i}}\lambda_{j_1}\lambda_{j_2}\cdots\lambda_{j_{m-1}}\;.
\end{eqnarray*}
Now let us show that the
second term of the last line of the above equation contains $m$ times of $\sigma_m$.
 For each specific term of $\sigma_m$, say
$\lambda_1\lambda_2\cdots\lambda_m$, it is included $m$ times in the
second term of the last line of the above equation, {\it i.e.}, it
is contained by the $m$ terms as follows
\begin{eqnarray*}
\lambda_i\cdot\sum_{\substack{j_1<j_2<\cdots<j_{m-1}\\j_l\neq
i}}\lambda_{j_1}\lambda_{j_2}\cdots\lambda_{j_{m-1}}\;,\qquad
i=1,2,\cdots,m\;.
\end{eqnarray*}
Thus,
\begin{eqnarray*}
\sigma_{m-1}T_1=\sum_{i=1}^n\lambda_i^2\cdot\sum_{\substack{j_1<j_2<\cdots<j_{m-2}\\j_l\neq
i}} \lambda_{j_1}\lambda_{j_2}\cdots\lambda_{j_{m-2}}+m\sigma_m\;.
\end{eqnarray*}
Next, multiplying $\sigma_{m-2}$ by
$(\lambda_1^2+\lambda_2^2+\cdots+\lambda_n^2)$ gives
\begin{eqnarray*}
\sigma_{m-2}T_2=\sum_{i=1}^n\lambda_i^3\cdot\sum_{\substack{j_1<j_2<\cdots<j_{m-3}\\j_l\neq
i}} \lambda_{j_1}\lambda_{j_2}\cdots\lambda_{j_{m-3}}
+\sum_{i=1}^n\lambda_i^2\cdot\sum_{\substack{j_1<j_2<\cdots<j_{m-2}\\j_l\neq
i}} \lambda_{j_1}\lambda_{j_2}\cdots\lambda_{j_{m-2}}\;.
\end{eqnarray*}
Likewise,
\begin{eqnarray*}
\sigma_{m-3}T_3&=&\sum_{i=1}^n\lambda_i^4\cdot\sum_{\substack{j_1<j_2<\cdots<j_{m-4}\\j_l\neq
i}} \lambda_{j_1}\lambda_{j_2}\cdots\lambda_{j_{m-4}}
+\sum_{i=1}^n\lambda_i^3\cdot\sum_{\substack{j_1<j_2<\cdots<j_{m-3}\\j_l\neq
i}} \lambda_{j_1}\lambda_{j_2}\cdots\lambda_{j_{m-3}}\;,\\
&\vdots&\\
\sigma_1T_{m-1}&=&\sum_{i=1}^n\lambda_i^m+\sum_{i=1}^n\lambda_i^{m-1}\sum_{j\neq
i}\lambda_j=T_m+\sum_{i=1}^n\lambda_i^{m-1}\sum_{j\neq
i}\lambda_j\;.
\end{eqnarray*}
Combing the above equations yields
\begin{eqnarray*}
\sigma_{m-1}T_1-\sigma_{m-2}T_2+\sigma_{m-3}T_3-\sigma_{m-4}T_4+\cdots+(-)^m\sigma_1T_{m-1}
=m\sigma_m+(-)^mT_m\;,
\end{eqnarray*}
which results in
\begin{eqnarray*}
mD_m+D_{m-1}T_1+D_{m-2}T_2+\cdots+D_1T_{m-1}+T_m=0\;.\qquad \Box
\end{eqnarray*}
\newline
From {\bfseries Theorem \ref{theorem1}}, the characteristic
polynomials for $n\times n$ matrices can be recursively obtained as
follows
\begin{eqnarray}
&&n=1\;,\qquad \chi_{\mathbf{A}}(\lambda)=\lambda-T_1\;,\nonumber\\
&&n=2\;,\qquad
\chi_{\mathbf{A}}(\lambda)=\lambda^2-T_1\lambda+\frac{1}{2}(T_1^2-T_2)\;,\\
&&n=3\;,\qquad
\chi_{\mathbf{A}}(\lambda)=\lambda^3-T_1\lambda^2+\frac{1}{2}(T_1^2-T_2)\lambda
-\frac{1}{6}(T_1^3-3T_1T_2+2T_3)\;,\nonumber\\
&&\cdots\cdots\nonumber
\end{eqnarray}
It is interesting to note that {\it the characteristic polynomial
coefficients $D_m$ ($1\leq m\leq n$) as polynomial functions of the
traces $T_k$ ($k=1,2,\ldots,n$) are formally unchanged when $n$
increases, though $T_k$ are obviously different for distinct $n$.}
As a consequence, {\it we can regard $D_m$ ($m=1,2,\ldots$) as an
infinite sequence.} {\bfseries Theorem \ref{theorem1}} gives
\begin{eqnarray}
D_1&=&-T_1\;,\nonumber\\
2D_2&=&-(T_1D_1+T_2)\;,\\
3D_3&=&-(T_1D_2+T_2D_1+T_3)\;,\nonumber\\
\cdots&&\cdots\nonumber
\end{eqnarray}
which leads to
\begin{eqnarray}
&&D_1+2xD_2+3x^2D_3+\cdots\nonumber\\
&=&-T_1-x(T_1D_1+T_2)-x^2(T_1D_2+T_2D_1+T_3)-\cdots\nonumber\\
&=&-\bigg[T_1(1+xD_1+x^2D_2+\cdots)+xT_2(1+xD_1+x^2D_2+\cdots)+\cdots\bigg]\;.\label{gf1}
\end{eqnarray}
If we define the generating functions $f(x)$ and $g(x)$ for the
infinite sequences $\{D_n\}$ and $\{T_{k+1}\}$ respectively as
follows
\begin{eqnarray}
f(x)\equiv \sum_{n=0}^\infty x^nD_n\quad (\mbox{with}\;\;D_0\equiv
1)\;,\qquad g(x)\equiv\sum_{k=0}^\infty x^{k}T_{k+1}\;,\label{gf2}
\end{eqnarray}
then
\begin{eqnarray}
f(0)=1\,,\quad
D_n=\frac{1}{n!}\frac{\mathrm{d}^n}{\mathrm{d}x^n}f(x)\bigg|_{x=0}\,,\quad
T_n=\frac{1}{(n-1)!}\frac{\mathrm{d}^{n-1}}{\mathrm{d}x^{n-1}}g(x)\bigg|_{x=0}\,,\quad
(n=1,2,\ldots)
\end{eqnarray}
and Eq. \eqref{gf1} implies
\begin{eqnarray}
\frac{\mathrm{d}}{\mathrm{d}x}f(x)=-g(x)f(x)\;,
\end{eqnarray}
or
\begin{eqnarray}
f(x)=\exp\bigg[-\int_0^xg(t)\mathrm{d}t\bigg]\;,
\end{eqnarray}
which gives
\begin{eqnarray}
f(x)&=&\exp\bigg[-\sum_{n=1}^\infty\frac{T_n}{n}x^n\bigg]
=\prod_{n=1}^\infty\exp\bigg[-\frac{T_n}{n}x^n\bigg]
=\prod_{n=1}^\infty\sum_{p_n=0}^\infty\frac{1}{p_n!}
\bigg[-\frac{T_n}{n}x^n\bigg]^{p_n}\nonumber\\
&=&\sum_{n=0}^\infty x^n\cdot\sum_{(p_1,p_2,\ldots,p_n)\in
\mathcal{S}_n}\,\prod_{m=1}^n\frac{1}{p_m!}\bigg[-\frac{T_m}{m}\bigg]^{p_m}\;,\label{gf-int}
\end{eqnarray}
where the set $\mathcal{S}_n$ is defined by all nonnegative-integer
solutions $\{(p_1,p_2,\ldots,p_n)\}$ of the equation
$p_1+2p_2+\cdots+np_n=n$. Comparing Eq. \eqref{gf-int} with Eq.
\eqref{gf2}, we obtain the expression for a generic term of the
sequence $\{D_n\}$ as follows
\begin{eqnarray}
D_n=\sum_{(p_1,p_2,\ldots,p_n)\in
\mathcal{S}_n}\,\prod_{m=1}^n\frac{1}{p_m!}\bigg[-\frac{T_m}{m}\bigg]^{p_m}\;.\label{Dn}
\end{eqnarray}
\newline
{\bfseries Example 1.} Consider the computation of $D_4$. All
nonnegative integer solutions  $(p_1, p_2, p_3, p_4)$ for the
equation $p_1+2p_2+3p_3+4p_4=4$ and their corresponding terms are
shown in Table \ref{D4}. Adding up all these terms gives
\begin{eqnarray}
D_4=-\frac{T_4}{4}+\frac{T_1T_3}{3}-\frac{T_1^2T_2}{4}+\frac{T_2^2}{8}+\frac{T_1^4}{24}\;.
\end{eqnarray}
\begin{table}[ht]
\caption{\label{D4}All nonnegative integer solutions $(p_1, p_2,
p_3, p_4)$ for the equation $p_1+2p_2+3p_3+4p_4=4$. For each
solution, there is a corresponding term
$\frac{1}{p_1!}(-T_1)^{p_1}\frac{1}{p_2!}(-\frac{T_2}{2})^{p_2}
\frac{1}{p_3!}(-\frac{T_3}{3})^{p_3}\frac{1}{p_4!}(-\frac{T_4}{4})^{p_4}$.
Summing of all the corresponding terms yields $D_4$.}
\begin{center}
\renewcommand{\arraystretch}{1.2}
\begin{tabular}{|c|c|c|c|c|}
\hline

~~$p_1$~~ & ~~$p_2$~~ & ~~$p_3$~~ & ~~$p_4$~~ & ~corresponding terms~ \\

\hline \hline

0 & 0 & 0 & 1 & $-\frac{T_4}{4}$ \\

\hline

1 & 0 & 1 & 0 & $(-T_1)(-\frac{T_3}{3})$ \\

\hline

2 & 1 & 0 & 0 & $\frac{1}{2!}(-T_1)^2(-\frac{T_2}{2})$ \\

\hline

0 & 2 & 0 & 0 & $\frac{1}{2!}(-\frac{T_2}{2})^2$ \\

\hline

4 & 0 & 0 & 0 & $\frac{1}{4!}(-T_1)^4$ \\

\hline
\end{tabular}
\end{center}
\end{table}
\newline
Now we state without proof the well-known theorem in linear algebra:
\begin{theorem}[Cayley-Hamilton]\label{theorem2}
Any $n\times n$ matrix $\mathbf{A}$ obeys its own characteristic
equation, that is, $\chi_{\mathbf{A}}(\mathbf{A})=\mathbf{0}$.
\end{theorem}

Combining Eq. \eqref{Dn} with {\bfseries Theorems \ref{theorem1}}
and {\bfseries \ref{theorem2}}, we obtain the following improved
results:
\begin{proposition}\label{prop1}
For any complex $n\times n$ matrix $\mathbf{A}$, its characteristic
polynomial is
$\chi_{\mathbf{A}}(\lambda)=\lambda^n+D_1\lambda^{n-1}+\cdots+D_{n-1}\lambda+D_n$,
and
$\chi_{\mathbf{A}}(\mathbf{A})=\mathbf{A}^n+D_1\mathbf{A}^{n-1}+\cdots
+D_{n-1}\mathbf{A}+D_n=\mathbf{0}$ where the coefficients are given
by
\begin{eqnarray}
D_k=\sum_{(p_1,p_2,\ldots,p_k)\in
\mathcal{S}_k}\,\prod_{m=1}^k\frac{1}{p_m!}\bigg[-\frac{T_m}{m}\bigg]^{p_m}\;,\qquad
(k=1,2,\cdots,n)
\end{eqnarray}
with $T_m$ standing for $\mathrm{tr}(\mathbf{A}^m)$ and
$\mathcal{S}_k$ being a set including all nonnegative-integer
solutions $\{(p_1,p_2,\ldots,p_k)\}$ of the equation
$p_1+2p_2+\cdots+kp_k=k$.
\end{proposition}

Note that the determinant of a matrix $\mathbf{A}$ is the product of
all the eigenvalues $\lambda_m$ ($m=1,2,\ldots,n$), that is,
$\det(\mathbf{A})=\sigma_n=\prod_{m=1}^n\lambda_m$. Thus, {\bfseries
Proposition \ref{prop1}} implies the following byproduct results:
\begin{proposition}
For any complex $n\times n$ matrix $\mathbf{A}$, its determinant is
given by
\begin{eqnarray}
\det(\mathbf{A})=(-)^nD_n=(-)^n\cdot\sum_{(p_1,p_2,\ldots,p_n)\in
\mathcal{S}_n}\,\prod_{m=1}^n\frac{1}{p_m!}\bigg[-\frac{T_m}{m}\bigg]^{p_m}\;,\label{det}
\end{eqnarray}
with $T_m$ standing for $\mathrm{tr}(\mathbf{A}^m)$ and
$\mathcal{S}_n$ being a set including all nonnegative-integer
solutions $\{(p_1,p_2,\ldots,p_n)\}$ of the equation
$p_1+2p_2+\cdots+np_n=n$.
\end{proposition}

\section{Applications to Chiral Perturbation Theory\label{sec-app-chpt}}

The trace formulae of the Cayley-Hamilton theorem can be used to
build relations between traces of finite dimensional matrices. Let
us study, case by case, the trace relations of $n\times n$ matrices
for $n=2$, 3, 4, and 5. In the following, $\langle\mathbf{A}\rangle$
is employed to denote the trace of a matrix $\mathbf{A}$.

\subsubsection{The $n=2$ case}

For any $2\times 2$ matrix $\mathbf{A}$, Cayley-Hamilton theorem
reads
\begin{eqnarray}
\mathbf{0}=\chi_{\mathbf{A}}(\mathbf{A})=\mathbf{A}^2-\langle\mathbf{A}\rangle\mathbf{A}
+\frac{1}{2}\big(\langle\mathbf{A}\rangle^2-\langle\mathbf{A}^2\rangle\big)\;.\label{CH-2}
\end{eqnarray}
Multiplying Eq. \eqref{CH-2} by $\mathbf{A}$ and taking the trace
results in
\begin{eqnarray}
0=F(\mathbf{A})\equiv\langle\mathbf{A}^3\rangle-\frac{3}{2}\langle\mathbf{A}^2\rangle
\langle\mathbf{A}\rangle+\frac{1}{2}\langle\mathbf{A}\rangle^3\;.\label{CH-2-1}
\end{eqnarray}
Substituting
$\mathbf{A}=\lambda_1\mathbf{A}_1+\lambda_2\mathbf{A}_2$, where
$\lambda_{1,2}$ are arbitrary parameters, into Eq. \eqref{CH-2-1}
one finds
\begin{eqnarray}
0&=&F(\lambda_1\mathbf{A}_1+\lambda_2\mathbf{A}_2)\nonumber\\
&=&\lambda_1^3F(\mathbf{A}_1)+\lambda_2^3F(\mathbf{A}_2)
+\lambda_1^2\lambda_2F_{12}(\mathbf{A}_1,\mathbf{A}_2)
+\lambda_2^2\lambda_1F_{21}(\mathbf{A}_1,\mathbf{A}_2)\nonumber\\
&=&\lambda_1^2\lambda_2F_{12}(\mathbf{A}_1,\mathbf{A}_2)
+\lambda_2^2\lambda_1F_{21}(\mathbf{A}_1,\mathbf{A}_2)\;.\label{CH-2-2}
\end{eqnarray}
Now, substituting
$\mathbf{A}=\lambda_1\mathbf{A}_1+\lambda_2\mathbf{A}_2+\lambda_3\mathbf{A}_3$
into Eq. \eqref{CH-2-1} leads to
\begin{eqnarray}
0&=&F(\lambda_1\mathbf{A}_1+\lambda_2\mathbf{A}_2+\lambda_3\mathbf{A}_3)\nonumber\\
&=&\sum_{i=1}^3\lambda_i^3F(\mathbf{A}_i)
+\sum_{i<j}\bigg[\lambda_i^2\lambda_jF_{ij}(\mathbf{A}_i,\mathbf{A}_j)
+\lambda_j^2\lambda_iF_{ji}(\mathbf{A}_i,\mathbf{A}_j)\bigg]
+\lambda_1\lambda_2\lambda_3F_{123}(\mathbf{A}_1,\mathbf{A}_2,\mathbf{A}_3)\nonumber\\
&=&\lambda_1\lambda_2\lambda_3F_{123}(\mathbf{A}_1,\mathbf{A}_2,\mathbf{A}_3)\;,
\end{eqnarray}
where the last equality comes from Eqs. \eqref{CH-2-1} and
\eqref{CH-2-2}. Thus,
$F_{123}(\mathbf{A}_1,\mathbf{A}_2,\mathbf{A}_3)=0$, that is,
\begin{eqnarray}
0=\sum_{2\mbox{\scriptsize~perm}}\langle\mathbf{A}_1\mathbf{A}_2\mathbf{A}_3\rangle
-\sum_{3\mbox{\scriptsize~perm}}\langle\mathbf{A}_1\mathbf{A}_2\rangle
\langle\mathbf{A}_3\rangle
+\langle\mathbf{A}_1\rangle\langle\mathbf{A}_2\rangle\langle\mathbf{A}_3\rangle\;.
\label{CH-2-c-3}
\end{eqnarray}

Likewise, multiplying Eq. \eqref{CH-2} by $\mathbf{A}^2$ and
subsequently taking the trace gives
\begin{eqnarray}
0=\langle\mathbf{A}^4\rangle-\langle\mathbf{A}^3\rangle\langle\mathbf{A}\rangle
-\frac{1}{2}\langle\mathbf{A}^2\rangle^2
+\frac{1}{2}\langle\mathbf{A}^2\rangle\langle\mathbf{A}\rangle^2\;.\label{CH-2-3}
\end{eqnarray}
Then inserting $\mathbf{A}=\sum_{i=1}^4\lambda_i\mathbf{A}_i$ into
Eq. \eqref{CH-2-3} and comparing the coefficients of
$\lambda_1\lambda_2\lambda_3\lambda_4$, one finds
\begin{eqnarray}
0=
\sum_{6\mbox{\scriptsize~perm}}\langle\mathbf{A}_1\mathbf{A}_2\mathbf{A}_3\mathbf{A}_4\rangle
-\frac{3}{4}\sum_{8\mbox{\scriptsize~perm}}\langle\mathbf{A}_1\mathbf{A}_2\mathbf{A}_3\rangle
\langle\mathbf{A}_4\rangle-\sum_{3\mbox{\scriptsize~perm}}\langle\mathbf{A}_1\mathbf{A}_2
\rangle\langle\mathbf{A}_3\mathbf{A}_4\rangle+\frac{1}{2}\sum_{6\mbox{\scriptsize~perm}}
\langle\mathbf{A}_1\mathbf{A}_2\rangle\langle\mathbf{A}_3\rangle\langle\mathbf{A}_4\rangle\;.
\label{CH-2-c-4}
\end{eqnarray}
In particular, if $\langle\mathbf{A}_i\rangle=0$, Eq.
\eqref{CH-2-c-4} becomes
\begin{eqnarray}
0=
\sum_{6\mbox{\scriptsize~perm}}\langle\mathbf{A}_1\mathbf{A}_2\mathbf{A}_3\mathbf{A}_4\rangle
-\sum_{3\mbox{\scriptsize~perm}}\langle\mathbf{A}_1\mathbf{A}_2
\rangle\langle\mathbf{A}_3\mathbf{A}_4\rangle\;,
\qquad(\mbox{for~}\langle\mathbf{A}_i\rangle=0)\;,\label{CH-2-c-4-1}
\end{eqnarray}
whose explicit form is
\begin{eqnarray}
&&\langle\mathbf{A}_1\mathbf{A}_2\mathbf{A}_3\mathbf{A}_4\rangle
+\langle\mathbf{A}_1\mathbf{A}_3\mathbf{A}_2\mathbf{A}_4\rangle
+\langle\mathbf{A}_2\mathbf{A}_1\mathbf{A}_3\mathbf{A}_4\rangle
+\langle\mathbf{A}_2\mathbf{A}_3\mathbf{A}_1\mathbf{A}_4\rangle
+\langle\mathbf{A}_3\mathbf{A}_1\mathbf{A}_2\mathbf{A}_4\rangle
+\langle\mathbf{A}_3\mathbf{A}_2\mathbf{A}_1\mathbf{A}_4\rangle\nonumber\\
&=&\langle\mathbf{A}_1\mathbf{A}_2\rangle\langle\mathbf{A}_3\mathbf{A}_4\rangle
+\langle\mathbf{A}_1\mathbf{A}_3\rangle\langle\mathbf{A}_2\mathbf{A}_4\rangle
+\langle\mathbf{A}_1\mathbf{A}_4\rangle\langle\mathbf{A}_2\mathbf{A}_3\rangle\;,
\qquad(\mbox{for~}\langle\mathbf{A}_i\rangle=0)\;.\label{CH-2-c-4-2}
\end{eqnarray}
Then further taking $\mathbf{A}_1=\mathbf{A}_2=\mathbf{A}$ and
$\mathbf{A}_3=\mathbf{A}_4=\mathbf{B}$ in Eq. \eqref{CH-2-c-4-2},
one obtains
\begin{eqnarray}
4\langle\mathbf{A}^2\mathbf{B}^2\rangle
+2\langle\mathbf{A}\mathbf{B}\mathbf{A}\mathbf{B}\rangle=
\langle\mathbf{A}^2\rangle\langle\mathbf{B}^2\rangle
+2\langle\mathbf{A}\mathbf{B}\rangle^2\;,
\qquad(\mbox{for~}\langle\mathbf{A}\rangle=\langle\mathbf{B}\rangle=0)\;.\label{CH-2-c-4-3}
\end{eqnarray}

Eventually, multiplying Eq. \eqref{CH-2} by $\mathbf{A}^4$ and the
same procedure as above results in
\begin{eqnarray}
0&=&
\sum_{120\mbox{\scriptsize~perm}}\langle\mathbf{A}_1\mathbf{A}_2\mathbf{A}_3\mathbf{A}_4
\mathbf{A}_5\mathbf{A}_6\rangle
-\frac{5}{6}\sum_{144\mbox{\scriptsize~perm}}\langle\mathbf{A}_1\mathbf{A}_2\mathbf{A}_3
\mathbf{A}_4\mathbf{A}_5\rangle\langle\mathbf{A}_6\rangle
-\frac{2}{3}\sum_{90\mbox{\scriptsize~perm}}\langle\mathbf{A}_1\mathbf{A}_2\mathbf{A}_3
\mathbf{A}_4\rangle\langle\mathbf{A}_5\mathbf{A}_6\rangle\nonumber\\
&&+\frac{2}{3}\sum_{90\mbox{\scriptsize~perm}}\langle\mathbf{A}_1\mathbf{A}_2\mathbf{A}_3
\mathbf{A}_4\rangle\langle\mathbf{A}_5\rangle\langle\mathbf{A}_6\rangle\;.\label{CH-2-c-6}
\end{eqnarray}
Eqs. \eqref{CH-2-c-3}, \eqref{CH-2-c-4} and \eqref{CH-2-c-6} are
useful relations for the construction of any 2-flavor chiral
perturbation theory.
\newline

{\bfseries Example 2.} Consider a 2-flavor chiral perturbation
theory with the chiral (global) symmetry breaking pattern
$[SU(2)_L\times SU(2)_R\times U(1)_V]/[SU(2)_V\times U(1)_V]$. The
basic building block of the theory is usually chosen to be a unitary
unimodular $2\times2$ matrix field $U$, which transforms as
$U\rightarrow RUL^\dag$ under the chiral symmetry group with $R\in
SU(2)_R$ and $L\in SU(2)_L$. If we define $X_\mu\equiv U^\dag(D_\mu
U)$ which transforms like $X_\mu\rightarrow LX_\mu L^\dag$. As $\det
U=1$, then
\begin{eqnarray}
0=\partial_\mu\det U=\partial_\mu(e^{\log\det U})=e^{\log\det
U}\partial_\mu(\log\det U)=\partial_\mu(\mathrm{tr}\log
U)=\mathrm{tr}(U^\dag\partial_\mu U)=\mathrm{tr}(U^\dag D_\mu
U)=\mathrm{tr}(X_\mu)\;.\label{traceless}
\end{eqnarray}
In the chiral Lagrangian, the chiral symmetric terms that consist of
four $X_\mu$'s are $\langle X_\mu X^\mu X_\nu X^\nu\rangle$,
$\langle X_\mu X_\nu X^\mu X^\nu\rangle$, $\langle X_\mu
X^\mu\rangle^2$, and $\langle X_\mu X_\nu\rangle^2$. Explicitly, the
first two terms are
\begin{eqnarray}
\langle X_\mu X^\mu X_\nu X^\nu\rangle&=&\sum_ig_{ii}^2\langle
X^iX^iX^iX^i\rangle +\sum_{i<k}g_{ii}g_{kk}\bigg(\langle
X^iX^iX^kX^k\rangle+\langle X^kX^kX^iX^i\rangle\bigg)\nonumber\\
&=&\sum_ig_{ii}^2\langle X^iX^iX^iX^i\rangle
+2\sum_{i<k}g_{ii}g_{kk}\langle X^iX^iX^kX^k\rangle\;,\\
\langle X_\mu X_\nu X^\mu X^\nu\rangle&=&\sum_ig_{ii}^2\langle
X^iX^iX^iX^i\rangle+2\sum_{i<k}g_{ii}g_{kk}\langle
X^iX^kX^iX^k\rangle\;.
\end{eqnarray}
Note that there is no summation on repeated indices $i$ and $k$
unless specially stated by a summation symbol, $\sum$. Then, a
linear combination of these two terms gives
\begin{eqnarray}
2\langle X_\mu X^\mu X_\nu X^\nu\rangle+\langle X_\mu X_\nu X^\mu
X^\nu\rangle&=&3\sum_ig_{ii}^2\langle X^iX^iX^iX^i\rangle
+\sum_{i<k}g_{ii}g_{kk}\bigg(4\langle X^iX^iX^kX^k\rangle+2\langle
X^iX^kX^iX^k\rangle\bigg)\nonumber\\
&=&\frac{3}{2}\sum_ig_{ii}^2\langle X^iX^i\rangle^2
+\sum_{i<k}g_{ii}g_{kk}\bigg(\langle X^iX^i\rangle\langle
X^kX^k\rangle+2\langle X^iX^k\rangle\langle X^iX^k\rangle\bigg)\nonumber\\
&=&\frac{1}{2}\langle X_\mu X^\mu\rangle^2+\langle X_\mu
X_\nu\rangle^2\;,\label{CH-2-example}
\end{eqnarray}
where the second equality comes from Eqs. \eqref{CH-2-3} and
\eqref{CH-2-c-4-3}. From Eq. \eqref{CH-2-example}, one finds that at
least one of the four 4-$X_\mu$ terms can be determined by the other
three terms. A check for Eq. \eqref{CH-2-example} by an alternative
method is given in App. \ref{check}.

\subsubsection{The $n=3$ case}

For any $3\times 3$ matrix $\mathbf{A}$, Cayley-Hamilton theorem
reads
\begin{eqnarray}
\mathbf{0}=\chi_{\mathbf{A}}(\mathbf{A})=\mathbf{A}^3-\langle\mathbf{A}\rangle\mathbf{A}^2
+\frac{1}{2}\big(\langle\mathbf{A}\rangle^2-\langle\mathbf{A}^2\rangle\big)\mathbf{A}
-\frac{1}{6}\big(\langle\mathbf{A}\rangle^3
-3\langle\mathbf{A}^2\rangle\langle\mathbf{A}\rangle
+2\langle\mathbf{A}^3\rangle\big)\;,\label{CH-3}
\end{eqnarray}
from which, the same procedure as the $n=2$ case yields
\begin{eqnarray}
0&=&
\sum_{6\mbox{\scriptsize~perm}}\langle\mathbf{A}_1\mathbf{A}_2\mathbf{A}_3\mathbf{A}_4\rangle
-\sum_{8\mbox{\scriptsize~perm}}\langle\mathbf{A}_1\mathbf{A}_2\mathbf{A}_3\rangle
\langle\mathbf{A}_4\rangle-\sum_{3\mbox{\scriptsize~perm}}\langle\mathbf{A}_1\mathbf{A}_2
\rangle\langle\mathbf{A}_3\mathbf{A}_4\rangle+\sum_{6\mbox{\scriptsize~perm}}
\langle\mathbf{A}_1\mathbf{A}_2\rangle\langle\mathbf{A}_3\rangle\langle\mathbf{A}_4\rangle
\nonumber\\
&&-\langle\mathbf{A}_1\rangle\langle\mathbf{A}_2\rangle\langle\mathbf{A}_3\rangle
\langle\mathbf{A}_4\rangle\;,\label{CH-3-c-4}\\
0&=&
\sum_{120\mbox{\scriptsize~perm}}\langle\mathbf{A}_1\mathbf{A}_2\mathbf{A}_3\mathbf{A}_4
\mathbf{A}_5\mathbf{A}_6\rangle
-\frac{5}{6}\sum_{144\mbox{\scriptsize~perm}}\langle\mathbf{A}_1\mathbf{A}_2\mathbf{A}_3
\mathbf{A}_4\mathbf{A}_5\rangle\langle\mathbf{A}_6\rangle
-\frac{2}{3}\sum_{90\mbox{\scriptsize~perm}}\langle\mathbf{A}_1\mathbf{A}_2\mathbf{A}_3
\mathbf{A}_4\rangle\langle\mathbf{A}_5\mathbf{A}_6\rangle\nonumber\\
&&+\frac{2}{3}\sum_{90\mbox{\scriptsize~perm}}\langle\mathbf{A}_1\mathbf{A}_2\mathbf{A}_3
\mathbf{A}_4\rangle\langle\mathbf{A}_5\rangle\langle\mathbf{A}_6\rangle
-\sum_{40\mbox{\scriptsize~perm}}\langle\mathbf{A}_1\mathbf{A}_2\mathbf{A}_3\rangle
\langle\mathbf{A}_4\mathbf{A}_5\mathbf{A}_6\rangle\nonumber\\
&&+\frac{1}{2}\sum_{120\mbox{\scriptsize~perm}}
\langle\mathbf{A}_1\mathbf{A}_2\mathbf{A}_3\rangle
\langle\mathbf{A}_4\mathbf{A}_5\rangle\langle\mathbf{A}_6\rangle
-\frac{1}{2}\sum_{40\mbox{\scriptsize~perm}}
\langle\mathbf{A}_1\mathbf{A}_2\mathbf{A}_3\rangle
\langle\mathbf{A}_4\rangle\langle\mathbf{A}_5\rangle\langle\mathbf{A}_6\rangle\;.
\label{CH-3-c-6}
\end{eqnarray}
Eqs. \eqref{CH-3-c-4} and \eqref{CH-3-c-6} are useful relations for
the construction of any 3-flavor chiral perturbation theory, such as
$[SU(3)_L\times SU(3)_R\times U(1)_V]/[SU(3)_V\times U(1)_V]$ and
$SU(3)/SO(3)$ chiral perturbation theories.

\subsubsection{The $n=4$ case}

For any $4\times 4$ matrix $\mathbf{A}$, Cayley-Hamilton theorem
reads
\begin{eqnarray}
\mathbf{0}=\chi_{\mathbf{A}}(\mathbf{A})&=&\mathbf{A}^4-\langle\mathbf{A}\rangle\mathbf{A}^3
+\frac{1}{2}\big(\langle\mathbf{A}\rangle^2-\langle\mathbf{A}^2\rangle\big)\mathbf{A}^2
-\frac{1}{6}\big(\langle\mathbf{A}\rangle^3
-3\langle\mathbf{A}^2\rangle\langle\mathbf{A}\rangle
+2\langle\mathbf{A}^3\rangle\big)\mathbf{A}\nonumber\\
&&+\bigg(-\frac{1}{4}\langle\mathbf{A}^4\rangle
+\frac{1}{3}\langle\mathbf{A}^3\rangle\langle\mathbf{A}\rangle
+\frac{1}{8}\langle\mathbf{A}^2\rangle^2
-\frac{1}{4}\langle\mathbf{A}^2\rangle\langle\mathbf{A}\rangle^2
+\frac{1}{24}\langle\mathbf{A}\rangle^4\bigg)\;,\label{CH-4}
\end{eqnarray}
from which, the same procedure as the $n=2$ case yields
\begin{eqnarray}
0&=&
\sum_{120\mbox{\scriptsize~perm}}\langle\mathbf{A}_1\mathbf{A}_2\mathbf{A}_3\mathbf{A}_4
\mathbf{A}_5\mathbf{A}_6\rangle
-\frac{5}{6}\sum_{144\mbox{\scriptsize~perm}}\langle\mathbf{A}_1\mathbf{A}_2\mathbf{A}_3
\mathbf{A}_4\mathbf{A}_5\rangle\langle\mathbf{A}_6\rangle
-\sum_{90\mbox{\scriptsize~perm}}\langle\mathbf{A}_1\mathbf{A}_2\mathbf{A}_3
\mathbf{A}_4\rangle\langle\mathbf{A}_5\mathbf{A}_6\rangle\nonumber\\
&&+\frac{2}{3}\sum_{90\mbox{\scriptsize~perm}}\langle\mathbf{A}_1\mathbf{A}_2\mathbf{A}_3
\mathbf{A}_4\rangle\langle\mathbf{A}_5\rangle\langle\mathbf{A}_6\rangle
-\sum_{40\mbox{\scriptsize~perm}}\langle\mathbf{A}_1\mathbf{A}_2\mathbf{A}_3\rangle
\langle\mathbf{A}_4\mathbf{A}_5\mathbf{A}_6\rangle\nonumber\\
&&+\frac{5}{6}\sum_{120\mbox{\scriptsize~perm}}
\langle\mathbf{A}_1\mathbf{A}_2\mathbf{A}_3\rangle
\langle\mathbf{A}_4\mathbf{A}_5\rangle\langle\mathbf{A}_6\rangle
-\frac{1}{2}\sum_{40\mbox{\scriptsize~perm}}
\langle\mathbf{A}_1\mathbf{A}_2\mathbf{A}_3\rangle
\langle\mathbf{A}_4\rangle\langle\mathbf{A}_5\rangle\langle\mathbf{A}_6\rangle\nonumber\\
&&+\sum_{15\mbox{\scriptsize~perm}}
\langle\mathbf{A}_1\mathbf{A}_2\rangle
\langle\mathbf{A}_3\mathbf{A}_4\rangle
\langle\mathbf{A}_5\mathbf{A}_6\rangle
-\frac{2}{3}\sum_{45\mbox{\scriptsize~perm}}
\langle\mathbf{A}_1\mathbf{A}_2\rangle
\langle\mathbf{A}_3\mathbf{A}_4\rangle
\langle\mathbf{A}_5\rangle\langle\mathbf{A}_6\rangle\nonumber\\
&&+\frac{1}{3}\sum_{15\mbox{\scriptsize~perm}}
\langle\mathbf{A}_1\mathbf{A}_2\rangle
\langle\mathbf{A}_3\rangle\langle\mathbf{A}_4\rangle
\langle\mathbf{A}_5\rangle\langle\mathbf{A}_6\rangle\;,
\end{eqnarray}
which is a useful relation for the construction of any 4-flavor
chiral perturbation theory, such as $[SU(4)_L\times SU(4)_R\times
U(1)_V]/[SU(4)_V\times U(1)_V]$, $SU(4)/SO(4)$, and $SU(4)/Sp(4)$
chiral perturbation theories.

\subsubsection{The $n=5$ case}

For any $5\times 5$ matrix $\mathbf{A}$, Cayley-Hamilton theorem
reads
\begin{eqnarray}
\mathbf{0}=\chi_{\mathbf{A}}(\mathbf{A})&=&\mathbf{A}^5-\langle\mathbf{A}\rangle\mathbf{A}^4
+\frac{1}{2}\big(\langle\mathbf{A}\rangle^2-\langle\mathbf{A}^2\rangle\big)\mathbf{A}^3
-\frac{1}{6}\big(\langle\mathbf{A}\rangle^3
-3\langle\mathbf{A}^2\rangle\langle\mathbf{A}\rangle
+2\langle\mathbf{A}^3\rangle\big)\mathbf{A}^2\nonumber\\
&&+\bigg(-\frac{1}{4}\langle\mathbf{A}^4\rangle
+\frac{1}{3}\langle\mathbf{A}^3\rangle\langle\mathbf{A}\rangle
+\frac{1}{8}\langle\mathbf{A}^2\rangle^2
-\frac{1}{4}\langle\mathbf{A}^2\rangle\langle\mathbf{A}\rangle^2
+\frac{1}{24}\langle\mathbf{A}\rangle^4\bigg)\mathbf{A}\nonumber\\
&&+\bigg(-\frac{1}{5}\langle\mathbf{A}^5\rangle
+\frac{1}{4}\langle\mathbf{A}^4\rangle\langle\mathbf{A}\rangle
+\frac{1}{6}\langle\mathbf{A}^3\rangle\langle\mathbf{A}^2\rangle
-\frac{1}{6}\langle\mathbf{A}^3\rangle\langle\mathbf{A}\rangle^2
-\frac{1}{8}\langle\mathbf{A}^2\rangle^2\langle\mathbf{A}\rangle\nonumber\\
&&+\frac{1}{12}\langle\mathbf{A}^2\rangle\langle\mathbf{A}\rangle^3
-\frac{1}{120}\langle\mathbf{A}\rangle^5\bigg)\;,\label{CH-5}
\end{eqnarray}
from which, the same procedure as the $n=2$ case yields
\begin{eqnarray}
0&=&
\sum_{120\mbox{\scriptsize~perm}}\langle\mathbf{A}_1\mathbf{A}_2\mathbf{A}_3\mathbf{A}_4
\mathbf{A}_5\mathbf{A}_6\rangle
-\sum_{144\mbox{\scriptsize~perm}}\langle\mathbf{A}_1\mathbf{A}_2\mathbf{A}_3
\mathbf{A}_4\mathbf{A}_5\rangle\langle\mathbf{A}_6\rangle
-\sum_{90\mbox{\scriptsize~perm}}\langle\mathbf{A}_1\mathbf{A}_2\mathbf{A}_3
\mathbf{A}_4\rangle\langle\mathbf{A}_5\mathbf{A}_6\rangle\nonumber\\
&&+\sum_{90\mbox{\scriptsize~perm}}\langle\mathbf{A}_1\mathbf{A}_2\mathbf{A}_3
\mathbf{A}_4\rangle\langle\mathbf{A}_5\rangle\langle\mathbf{A}_6\rangle
-\sum_{40\mbox{\scriptsize~perm}}\langle\mathbf{A}_1\mathbf{A}_2\mathbf{A}_3\rangle
\langle\mathbf{A}_4\mathbf{A}_5\mathbf{A}_6\rangle\nonumber\\
&&+\sum_{120\mbox{\scriptsize~perm}}
\langle\mathbf{A}_1\mathbf{A}_2\mathbf{A}_3\rangle
\langle\mathbf{A}_4\mathbf{A}_5\rangle\langle\mathbf{A}_6\rangle
-\sum_{40\mbox{\scriptsize~perm}}
\langle\mathbf{A}_1\mathbf{A}_2\mathbf{A}_3\rangle
\langle\mathbf{A}_4\rangle\langle\mathbf{A}_5\rangle\langle\mathbf{A}_6\rangle\nonumber\\
&&+\sum_{15\mbox{\scriptsize~perm}}
\langle\mathbf{A}_1\mathbf{A}_2\rangle
\langle\mathbf{A}_3\mathbf{A}_4\rangle
\langle\mathbf{A}_5\mathbf{A}_6\rangle
-\sum_{45\mbox{\scriptsize~perm}}
\langle\mathbf{A}_1\mathbf{A}_2\rangle
\langle\mathbf{A}_3\mathbf{A}_4\rangle
\langle\mathbf{A}_5\rangle\langle\mathbf{A}_6\rangle\nonumber\\
&&+\sum_{15\mbox{\scriptsize~perm}}
\langle\mathbf{A}_1\mathbf{A}_2\rangle
\langle\mathbf{A}_3\rangle\langle\mathbf{A}_4\rangle
\langle\mathbf{A}_5\rangle\langle\mathbf{A}_6\rangle
-\langle\mathbf{A}_1\rangle\langle\mathbf{A}_2\rangle
\langle\mathbf{A}_3\rangle\langle\mathbf{A}_4\rangle
\langle\mathbf{A}_5\rangle\langle\mathbf{A}_6\rangle\;,
\end{eqnarray}
which is a useful relation for the construction of any 5-flavor
chiral perturbation theory, such as $[SU(5)_L\times SU(5)_R\times
U(1)_V]/[SU(5)_V\times U(1)_V]$ and $SU(5)/SO(5)$ chiral
perturbation theories.

\section{Applications to General Relativity\label{sec-gr}}

In general relativity, a second-rank covariant tensor field
$g_{\mu\nu}$ called the metric plays a crucial role. Its
determinant, $g\equiv\det(g_{\mu\nu})$, is usually needed to be
figured out in case one wants to handle the pseudoscalar volumn
element: $\sqrt{-g}d^dx$, where the superscript $d$ stands for the
number of spacetime dimensions. Let the dimension number $d$ be 10,
which is required for the consistency of superstring theories. Using
Eq. \eqref{det}, one can express the determinant of the metric
$10\times10$ matrix $\mathbf{g}$ in terms of the traces of the
successive powers of $\mathbf{g}$. Since there are 42 solutions
$(p_1,p_2,\ldots,p_{10})$ of the equation
$p_1+2p_2+\cdots+10p_{10}=10$, there are also 42 terms in the
expression of $\det\mathbf{g}$, which is given by
\begin{eqnarray}
\det(\mathbf{g})&=&-\frac{1}{10}T_{10}+\frac{1}{9}T_9T_1+\frac{1}{16}T_8T_2
-\frac{1}{16}T_8T_1^2
+\frac{1}{21}T_7T_3-\frac{1}{14}T_7T_2T_1+\frac{1}{42}T_7T_1^3+\frac{1}{24}T_6T_4
-\frac{1}{18}T_6T_3T_1\nonumber\\
&&-\frac{1}{48}T_6T_2^2+\frac{1}{24}T_6T_2T_1^2-\frac{1}{144}T_6T_1^4+\frac{1}{50}T_5^2
-\frac{1}{20}T_5T_4T_1-\frac{1}{30}T_5T_3T_2+\frac{1}{30}T_5T_3T_1^2
+\frac{1}{40}T_5T_2^2T_1\nonumber\\
&&-\frac{1}{60}T_5T_2T_1^3+\frac{1}{600}T_5T_1^5-\frac{1}{64}T_4^2T_2+\frac{1}{64}T_4^2T_1^2
-\frac{1}{72}T_4T_3^2+\frac{1}{24}T_4T_3T_2T_1-\frac{1}{72}T_4T_3T_1^3+\frac{1}{192}T_4T_2^3
\nonumber\\
&&-\frac{1}{64}T_4T_2^2T_1^2+\frac{1}{192}T_4T_2T_1^4-\frac{1}{2880}T_4T_1^6
+\frac{1}{162}T_3^3T_1+\frac{1}{144}T_3^2T_2^2-\frac{1}{72}T_3^2T_2T_1^2
+\frac{1}{432}T_3^2T_1^4\nonumber\\
&&-\frac{1}{36}T_3T_2^3T_1+\frac{1}{144}T_3T_2^2T_1^3-\frac{1}{720}T_3T_2T_1^5
+\frac{1}{15120}T_3T_1^7-\frac{1}{240}T_2^5+\frac{1}{96}T_2^4T_1^2\nonumber\\
&&-\frac{1}{288}T_2^3T_1^4+\frac{1}{5760}T_2^2T_1^6-\frac{1}{80640}T_2T_1^8
+\frac{1}{3628800}T_1^{10}\;,\label{det-g-10}
\end{eqnarray}
where $T_m$ ($m=1,2,\ldots,10$) stand for
$\mathrm{tr}(\mathbf{g}^m)$. Eq. \eqref{det-g-10} shows that the
determinant $g=\det(\mathbf{g})$ is completely decided by the 10
traces, $T_m$ ($m=1,2,\ldots,10$), and this expression is fairly
controllable and thus can be possibly used in future theoretical
considerations. By using {\it PASCAL} (a sample code is given in
App. \ref{pascal-code}), one easily finds that there are 56 trace
terms among $\det(\mathbf{g})$ for 11 dimensional spacetime
corresponding to the M theory, while 2436 terms for 26 dimensional
spacetime.

\section{Conclusions\label{sec-con}}
In this paper, we have given a new proof, by using combinatorics,
for the recurrence relations of the characteristic polynomial
coefficients, and, moreover, we have obtained an explicit expression
for the generic term of the coefficient sequence, which yields an
improved version of the Cayley-Hamilton's theorem with all
coefficients explicitly given, and which implies a byproduct
expression for the determinant of any finite-dimensional matrix in
terms of the traces of its successive powers. We have shown that
these relations can be applied to chiral perturbation theories and
general relativity. On the one hand, we have obtained the
Cayley-Hamilton relations up to the trace of six operators in any
$n$-flavor chiral perturbation theory for $n$ runs from 2 to 5. On
the other hand, we have obtained the determinant of the metric
tensor in terms of the traces of its successive powers in 10
dimensional spacetime. For higher dimensional spacetime, it is also
easily obtained by using a {\it PASCAL} program code, which is
contained in the appendix.

\appendix

\section{A Check for the Cayley-Hamilton Relations in the $n=2$ Case\label{check}}

In this appendix, we give a check for the Cayley-Hamilton relations
in the $n=2$ case by using the property of Pauli matrices. For
definiteness, let us verify Eq. \eqref{CH-2-example} in {\bfseries
Example 2}. Since $X_\mu$ is traceless due to Eq. \eqref{traceless},
then it can be decomposed, based on Pauli matrices, to
$X_\mu=\tau^aX_\mu^a$ with the coefficients $X_\mu^a$ $(a=1,2,3)$
and $\tau^a$ $(a=1,2,3)$ standing for Pauli matrices. Now let us
show the following identity:
\begin{eqnarray}
\langle X_\mu X_\nu X_\rho X_\sigma\rangle=\frac{1}{2}\bigg(\langle
X_\mu X_\nu\rangle\langle X_\rho X_\sigma\rangle+\langle X_\mu
X_\sigma\rangle\langle X_\nu X_\rho\rangle-\langle X_\mu
X_\rho\rangle\langle X_\nu X_\sigma\rangle\bigg)\;,\label{App-4X}
\end{eqnarray}
where $\langle\cdots\rangle$ stands for taking the trace.

{\bfseries Proof.} The well-known identities
$\tau^a\tau^b=\delta^{ab}+i\epsilon^{abc}\tau^c$ and
$\epsilon^{abe}\epsilon^{cde}=\delta^{ac}\delta^{bd}-\delta^{ad}\delta^{bc}$
imply that
\begin{eqnarray*}
\langle\tau^a\tau^b\rangle=2\delta^{ab}\;,\qquad
\langle\tau^a\tau^b\tau^c\tau^d\rangle=2(\delta^{ab}\delta^{cd}+\delta^{ad}\delta^{bc}
-\delta^{ac}\delta^{bd})\;.
\end{eqnarray*}
Thus,
\begin{eqnarray*}
\langle X_\mu X_\nu X_\rho
X_\sigma\rangle&=&\langle\tau^a\tau^b\tau^c\tau^d\rangle
X_\mu^aX_\nu^bX_\rho^cX_\sigma^d\\
&=&2(\delta^{ab}\delta^{cd}+\delta^{ad}\delta^{bc}
-\delta^{ac}\delta^{bd})X_\mu^aX_\nu^bX_\rho^cX_\sigma^d\\
&=&\frac{1}{2}\bigg(\langle X_\mu X_\nu\rangle\langle X_\rho
X_\sigma\rangle+\langle X_\mu X_\sigma\rangle\langle X_\nu
X_\rho\rangle-\langle X_\mu X_\rho\rangle\langle X_\nu
X_\sigma\rangle\bigg)\;.\qquad \Box
\end{eqnarray*}

Then, by contracting indices in Eq. \eqref{App-4X}, we obtain
\begin{eqnarray}
&&\langle X_\mu X_\nu X^\mu X^\nu\rangle=\langle X_\mu
X_\nu\rangle^2-\frac{1}{2}\langle X_\mu X^\mu\rangle^2\;,\label{App-4X-1}\\
&&\langle X_\mu X^\mu X_\nu X^\nu\rangle=\frac{1}{2}\langle X_\mu
X^\mu\rangle^2\;,\label{App-4X-2}
\end{eqnarray}
from which it is easily checked that
\begin{eqnarray}
2\langle X_\mu X^\mu X_\nu X^\nu\rangle+\langle X_\mu X_\nu X^\mu
X^\nu\rangle=\frac{1}{2}\langle X_\mu X^\mu\rangle^2+\langle X_\mu
X_\nu\rangle^2\;,\label{App-4X-CH}
\end{eqnarray}
which is exactly Eq. \eqref{CH-2-example} obtained in {\bfseries
Example 2}. It should be kept in mind that Eq. \eqref{App-4X}, or
Eqs. \eqref{App-4X-1} and \eqref{App-4X-2} hold only for $n=2$,
though they are stronger than Eq. \eqref{App-4X-CH} which comes from
the Cayley-Hamilton theorem. However, from Eq. \eqref{CH-3-c-4}, one
notes that the following relation holds as well in the $n=3$ case.
\begin{eqnarray}
0=
\sum_{6\mbox{\scriptsize~perm}}\langle\mathbf{A}_1\mathbf{A}_2\mathbf{A}_3\mathbf{A}_4\rangle
-\sum_{3\mbox{\scriptsize~perm}}\langle\mathbf{A}_1\mathbf{A}_2
\rangle\langle\mathbf{A}_3\mathbf{A}_4\rangle\;,
\qquad(\mbox{for~}\langle\mathbf{A}_i\rangle=0)\;.
\end{eqnarray}
As a consequence, Eq. \eqref{App-4X-CH} holds not only for
$2\times2$ but also for $3\times3$ traceless matrices $X_\mu$.
Obviously, this relation is invalid for $n\times n$ ($n\geq 4$)
matrices.

\section{A PASCAL Code for Computation of $D_n$\label{pascal-code}}
\begin{verbatim}
Type splittype=array[1..40]of integer;
var num,sum,k:integer;
var spl:splittype;
var f1,f2:text;
procedure Output;
  var i,x,y:integer;
  var sp:splittype;
  begin
    for i:=1 to 40 do sp[i]:=0;
    for i:=1 to k-2 do
      begin
        write(f1,spl[i],'+');
        sp[spl[i]]:=sp[spl[i]]+1;
      end;
    write(f1,spl[k-1],'=',num);sp[spl[k-1]]:=sp[spl[k-1]]+1;
    writeln(f1);
    for i:=1 to num do write(f2,sp[i],' ');
    i:=1;
    while sp[i]=0 do i:=i+1;
    if sp[i]<>1 then write(f2,'1/',sp[i],'!','(-T_',i,'/',i,')^',sp[i])
      else write(f2,'(-T_',i,'/',i,')');
    i:=i+1;
    while i<=num do
      begin
        if sp[i]<>0 then
          if sp[i]=1 then write(f2,'*(-T_',i,'/',i,')')
            else write(f2,'*1/',sp[i],'!','(-T_',i,'/',i,')^',sp[i],'*');
        i:=i+1;
      end;
    writeln(f2);
  end;
procedure split(n,m:integer);
  var i:integer;
  begin
    if n=0 then begin Output; sum:=sum+1 end
      else if n>0 then
        begin
          for i:=m downto 1 do
            begin
              if n>=i then spl[k]:=i;
              k:=k+1;
              split(n-i,i);
              k:=k-1;
            end;
        end;
  end;
begin
  sum:=0;k:=1;
  assign(f1,'Output.txt');
  assign(f2,'Output2.txt');
  rewrite(f1);
  rewrite(f2);
  readln(num);
  split(num,num);
  writeln(f1,'Total is ',sum);
  writeln(f2,'Total is ',sum);
  close(f1);
  close(f2);
end.
\end{verbatim}

\begin{acknowledgments}
We would like to thank J.~K.~Parry, Zhan~Xu and Ran~Lu for helpful
discussions. This work is supported in part by the National Natural
Science Foundation of China.
\end{acknowledgments}

\end{document}